\documentclass[%
prl
,twocolumn%
,secnumarabic%
,superscriptaddress%
,amssymb, amsmath,nobibnotes, aps, prl, footinbib]{revtex4}

\usepackage{graphicx}
\usepackage{bm}

\begin{document}

\title{The Effects of the Sagittarius Dwarf Tidal Stream on Dark
  Matter Detectors}

\author{Katherine Freese}
\affiliation{Michigan Center for Theoretical Physics, University of Michigan, 
  Ann Arbor, MI 48109}

\author{Paolo Gondolo}
\affiliation{Dept.\ of Physics, University of Utah, 115 S 1400 E \#201, 
  Salt Lake City, UT 84112}

\author{Heidi Jo Newberg}
\affiliation{Rensselaer Polytechnic Institute, Dept. of Physics, Troy, NY 12180}

\author{Matthew Lewis}
\affiliation{Michigan Center for Theoretical Physics, University of Michigan, 
  Ann Arbor, MI 48109}

\begin{abstract} 
\noindent
The Sagittarius dwarf tidal stream may be showering dark matter onto the solar neighborhood, which can change the results and interpretation of WIMP direct detection experiments. Stars in the stream may already have been detected in the solar neighborhood, and the dark matter in the stream is (0.3-25)\% of the local density. Experiments should see an annually modulated steplike feature in the energy recoil spectrum that would be a smoking gun for WIMP detection. The total count rate in detectors is not a cosine curve in time and peaks at a different time of year than the standard case.

\end{abstract}

\date{December 11, 2003}%
\maketitle

Dark matter in the halo of our galaxy may consist of WIMPs (weakly
interacting massive particles)\@.  More than twenty collaborations
worldwide are developing detectors designed to search for WIMPs through their elastic scattering off nuclei.  The
most frequently assumed velocity distribution for galactic dark matter
is that of a simple isothermal sphere.  However, recent observations
of the stellar component of the Galactic halo show evidence of a
merger history that has not yet become well mixed, and indicate that
halos form hierarchically.  The Sloan Digital Sky Survey and the Two
Micron All Sky Survey \cite{newberg,majewski} have traced the tidal
stream \cite{sagstr} of the Sagittarius dwarf galaxy (Sgr), a
satellite galaxy that is located inside the Milky Way Galaxy.  Two
streams of matter are being tidally pulled away from the main body of
the Sgr galaxy and extend outward from it.  The leading tail is
showering matter down upon the solar neighborhood \cite{majewski}.  Here we
show the effect of this stream on WIMP dark matter direct detection
experiments.
Further details can be found in
our longer paper \cite{longer}.

Estimates of the current mass-to-light ratio of the Sgr dwarf (and
hence plausibly the tidal tails) are $M/L \sim (25-100)$
\cite{majewski,ibata97}.  We estimate a range for the local stream
dark matter density \cite{longer}:
$ 
\rho_{str} = [2, 190] \times 10^4 
M_\odot/{\rm kpc}^3 = [0.001, 0.07] \, {\rm GeV/cm^3} ,
$ 
which corresponds to (0.3-25)\% of the local density of the isothermal
Galactic halo, assuming $\rho_h = 0.3$GeV/cm$^3$\footnote{
  Based on hierarchical clustering, it is
  likely that a clump contributes 1-5\% to the local dark
  matter density \protect\cite{swf}.}.
At the solar position, we expect the tidal stream to travel
roughly orthogonal to the Galactic plane with a speed of $\sim 300$
km/s.  We will suggest an identification of the Sgr stream with a
previously discovered \cite{helmi} coherently moving group of stars.
Other (lower density) tidal streams may pass by the Solar neighborhood too.

The additional WIMP flux from the Sgr stream gives a 0.3--25\%
increase in the count rate in detectors at energies below a
characteristic energy $E_c$, the highest energy that stream WIMPs can
impart to a target nucleus.  Thus, as pointed out by
\cite{gondologelmini,swf} for generic WIMP streams, we predict a
step-like feature in the energy recoil spectrum.

The count rate in WIMP detectors will experience an annual modulation
of amplitude $\sim 10\%$ due to Earth's motion around the Sun
\cite{dfs, ffg}.  For an isothermal halo, the peak count rate is on
June 2. The count rate of the Sgr stream is also annually modulated,
with a peak date different from that of the halo. Hence, as pointed
out by \cite{gondologelmini,swf} for generic streams, the total signal
(from halo plus stream) has a modified time dependence and peak date.
The step energy $E_c$ is also annually modulated; detection of this
modulation would provide powerful confirmation of a WIMP stream
signal.  For WIMP masses heavier than 50 GeV, the step lies above the
threshold energy of current and upcoming dark matter experiments, and
would affect their results. If their signal is due to WIMPs, DAMA
\cite{DAMA} should have the stream in their current data.

{\it Local stars in the Sagittarius tidal tails.--} Seven of 97 stars
(7\%) identified as halo stars by Helmi {\it et al} \cite{helmi}, within
one kpc of the Sun, are moving coherently in one direction.  Recently,
Majewski et al.\ \cite{majewski} noticed that the density of this
clump of giant stars in the solar region is similar to the expected
local density of stars from the Sgr dwarf tidal stream.  However, the
$Y-$component of the stars' velocity is too high to be consistent with
the Sgr dwarf orbital plane.  The velocity vector of these stars is
fifteen degrees away from the orbital plane of the Sgr dwarf
spheroidal galaxy, as measured by the positions of the stars in the
tidal tails, consistent with a plane that is tipped further from the
Galactic pole.

In order to determine their velocity components, we re-extracted 8
clump stars from Chiba and Yoshii \cite{chibayoshii}.  We confirmed
Helmi {\it et al}'s velocity of $290 \pm 26$ km/s and found that the direction 
of the stars' motion is $(l,b) = (116, -59)$.  In Galactic
coordinates\footnote{We take $\hat{X}$ in the direction of the
  Galactic center, so the Sun is at $(X,Y,Z) = (-8.5, 0, 0)$.}, the
mean velocity of the local stellar clump is $(V_X, V_Y, V_Z) = (-65
\pm 22, 135 \pm 12, -249 \pm 6)$, with velocity dispersions
$(\sigma_{V_X}, \sigma_{V_Y}, \sigma_{V_Z}) = (62, 33, 17)$ km/s.  Our
measurement of the dispersion in the $X$ direction is significantly
smaller than previously measured \cite{helmi}.  Since the experimental
errors in the space velocities of these stars are of the order of 10
km/s in each component, the asymmetric velocity dispersions are likely
to be inherent to this kinematic component, and not a result of
measurement error.

Except for the fact that the Helmi et al.\ stars are in a less polar
orbit than the Sgr dwarf, the characteristics are as expected for the
stars associated with the Sgr dwarf spheroidal galaxy.  They are a
spatially distinct component (surveying larger volumes does not
produce a star count that increases linearly with the volume)
\cite{chibabeers}.  They have very nearly the predicted speed
\cite{longer} and expected stellar density \cite{majewski}.  They are
the most significant low-velocity-dispersion component at the solar
position.

Given the large number of coincidences with our expectations for 
Sgr stream stars, it is logical to revisit whether a velocity
direction that differs from the Sgr orbital plane by $15^\circ$ rules
out an association.  The $Y-$component of the Helmi et al.\ stars is
5$\sigma$ higher than that required to match the published Sgr plane.
However, it is conceivable that the Sgr stream properties in the solar
neighborhood deviate from the Sgr dwarf orbital plane.

In Figure 5 of Newberg et al.~\cite{newberg}, it is shown that the
known tidal tails of the Sgr dwarf galaxy do not exactly follow the
Sgr dwarf orbital plane as fit to M-giants \cite{majewski}.  Fitting a
plane to the positions of a piece of the leading tail debris yields a
plane that is discrepant by $8^\circ$ from the nominal orbital plane.
In \cite{longer}, we estimated a stream width of 6 kpc for a piece of
the Sgr tidal tail that was 34 kpc from the Galactic center.  Though
the center of the tidal stream could in principle remain in one plane,
the individual stars in a stream with this width must differ in
orbital plane by $10^\circ$ or more from one side of the stream to the
other. A spatial separation of the velocity components is likely,
given that the stars are primarily stripped at the dwarf galaxy's
closest approach to the Galactic center \cite{johnston}, and are not
cast off along the length of the orbit.

Additional contributions to the orbital velocities of stars in a
restricted portion of the tidal stream include a non-spherical
component of the Galactic potential, any halo dark matter ``lumps"
\cite{lumpyhalo} yet to be discovered, and any rotation of the parent
galaxy to the present-day Sgr dwarf spheroidal.  All of these unknowns
together make it plausible that a local portion of the Sgr tidal
stream could have different orbital parameters than naively expected
from a global fit to the tidal tails.

Once stripped from the dwarf, stars and dark matter released into the
Galactic potential at the same position and with the same velocity
travel together.  The Helmi et al.\ stars provide the most likely
velocities for a tidal stream that passes through our location in the
Galaxy (though if it is not the Sgr stream, then the dark matter
density associated with it is currently unknown).  In the remainder of
this letter, we explore the implications of this tidal stream for WIMP
direct detection experiments, assuming it has the dark matter density
estimated for the Sgr tidal stream.

{\it Count Rates:} In WIMP direct detection experiments, the
differential detection rate per unit detector mass (counts/day/kg
detector/keV recoil energy) is \cite{gondologelmini}
\begin{equation}
\label{eq:rate}
  \frac{dR}{dE} = \frac{ \sigma_0 F^2(q) } { 2 m \mu^2 } 
\bigl[ \rho_h \eta_{h}(E,t) + \rho_{\rm str} \eta_{\rm str}(E,t) \bigr]
\end{equation}
where 
$\eta_h$ and $\eta_{str}$ are the mean
inverse speed of WIMPs in the standard Galactic halo and the stream
respectively,
$m$ is the WIMP mass, $M$ is the target nucleus mass,
$\mu \equiv m M/ (m +
M)$ is the reduced mass, $\sigma_0$ is the total nucleus-WIMP
cross section, $q = \sqrt{2 M E}$ is the nucleus recoil momentum, and
$F(q)$ is a nuclear form factor.  More detail can be found in our
longer paper \cite{longer}.  We consider two different sources of
WIMPs that contribute to count rates in detectors: WIMPs in the Milky
Way halo and WIMPs in the Sgr stream.  For the halo WIMPs, we assume
an isothermal sphere.  The Galactic WIMP speeds obey a Maxwellian
distribution with a velocity dispersion $\sigma_h$ truncated at the
escape velocity $v_{\rm esc}$.  We take $\sigma_h = 270$ km/s and
$v_{\rm esc} = 650$ km/s.  The resultant rates can be found in
\cite{ffg,longer}.

{\it Sgr Stream Component:} 
As our reference case, we use a WIMP stream velocity ${\bf
v}_{\rm str} = (-65,135,-249)\,\, {\rm km/s} \, .$ 
The characteristic
energy of the step in the recoil spectrum is
\begin{equation}
  E_{c}(t) = (2 \mu^2 /M) | {\bf v}_{\rm D}(t) -
{\bf v}_{\rm str} |^2 ,
\end{equation}
where ${\bf v}_{\rm D}(t)$ is the detector velocity.
This step energy is the maximum recoil energy that can be imparted to
the nucleus. The maximum momentum transferred from a WIMP to a nucleus
occurs when the WIMP bounces back and is $ q_{\rm max} = 2 \, \mu \,
\left| {\bf v}_{\rm D}(t) - {\bf v}_{\rm str} \right| $; the maximum
nuclear recoil energy then follows as $E_c(t) = q_{\rm max}^2/(2M)$.

The effect of a velocity dispersion $\sigma_{\rm str}$ in the Sgr
stream is to smooth out the edges of the step.  We assume that the
WIMPs in the Sgr stream follow a Maxwellian velocity distribution 
$ f_{\rm str}({\bf v}) = 
{(2\pi\sigma_{\rm str}^2/3)^{-3/2}} e^{-3|{\bf v}-{\bf v}_{\rm
    str}|^2/2\sigma_{\rm str}^2}$.
We will use $\sigma_{\rm str}
=$ 72 km/sec, but will study the effects of an
anisotropic $\sigma_{\rm str}$ in a later paper.

{\it Sgr Stream Annual Modulation:} 
The position of the step depends on the time of year:
\begin{equation}
  E_c(t) = E_c^{(0)} \left\{ 1 + A_c \cos[ \omega (t - t_c)] \right\} ,
\end{equation}
with $ \omega = 2 \pi/1$~yr, and $ E_c^{(0)}$ and $ A_c$ given in
\cite{longer}.  For the Sgr stream, we find that the amplitude of the
annual modulation of the step energy is $A_c= 25.2\%$.  The location
of the step is at the highest energy $E_{c,max}$ on Jan.\ 15 and
lowest energy $E_{c,min}$ on July 16.  The amplitude and phase of the
$E_c$ modulation do not depend on the target nucleus, WIMP mass, or
density of stream WIMPs, but do depend on the size and direction of
the stream velocity.

The count rates from both the halo and stream WIMPs also experience
annual modulation.  Since the halo and stream count rates peak on
different days of the year, the peak date of the total signal depends
on the fractional stream contribution.
%
The peak date of the stream contribution depends on the stream
direction and the recoil energy.  For our reference direction, at
recoil energies below $E_{c,min}$, the stream count rate (dominated at
low recoil energies by slow WIMPs relative to the Earth) peaks on July
16 and is minimum on Jan.\ 15.  Above $E_{c,min}$,
fast stream WIMPs (relative to the Earth) dominate the stream count
rate, and the phase of the annual modulation reverses: the stream
count rate peaks on Jan.\ 15 and is minimum on July 16.
For recoil energies above $E_{c,max}$,
the stream disappears entirely, and one returns to the annual
modulation of the halo, with a peak on June 2.  The phase of the
modulation of $E_c$ is 180$^\circ$ out of phase with the stream
modulation below $E_{c,min}$, and in phase with the stream modulation
between $E_{c,min}$ and $E_{c,max}$ (above which the stream signal
disappears).

At a given energy recoil, the total observed signal can have a variety
of periodicities that do not resemble sinusoids over the course of a
year.  For example, the peak of the stream signal on Jan.  15 can
produce a bump in the time series of the total signal.  Since the
stream peak depends sensitively on the direction of the stream, the
total count rate of halo plus stream can vary widely depending on the
stream contribution.  In particular, the peak date of the total
observable signal can vary over the entire year, depending on the
stream contribution.  Any stream component can thus drastically affect
what experimenters will find.

\begin{table}[tbp]
\begin{tabular}{|l||r|r||r|r||r|r|}
\hline
Target &
  \multicolumn{2}{c}{$m$=60 GeV} \vline &
    \multicolumn{2}{c}{$m$=100 GeV} \vline & 
      \multicolumn{2}{c}{$m$=500 GeV} \vline \\
\cline{2-7}
nucleus& 
  $E_{c,min}$ &$E_{c,max}$ & $E_{c,min}$ &$E_{c,max}$ &
      $E_{c,min}$ &$E_{c,max}$ \\
\hline
${}^{127}$I &
  1.87  & 2.50 &
    3.47 & 4.64 &
      10.8 & 14.5 \\
\hline
${}^{23}$Na &
  5.43 & 7.26 &
    6.79 & 9.08 &
      9.22 & 12.3 \\
\hline
\end{tabular}
\caption{Step energy $E_c$ in keVee
from WIMPs in the Sgr stream for Na and I on maximum and minimum
dates, Jan.\ 15 and July 16, for our reference case.  }\end{table}

{\it Results:}
As an illustrative example, we here focus on the NaI detector of the
DAMA experiment.  We assume $\sigma_0 = 7.2 \times 10^{-42} {\rm ~cm}^2 (\mu/\mu_p)^2 A^2$, where $A$ is the atomic mass and $\mu_p$ is the reduced WIMP-proton mass.
As discussed in \cite{longer}, the stream should be
visible in other experiments as well. 

DAMA has an energy threshold of 2 keV electron
equivalent (keVee), which corresponds to 22 keV recoil energy for
iodine \footnote{The conversion from keVee to keV recoil energy is
found from the quenching factor, 0.09 for I and 0.3
for Na.}.   In Table I, we list the values
of the step energy $E_c$ due to Sgr stream WIMPs for our reference
case \footnote{For DAMA, there are two steps, at
  lower energy due to I and at higher energy due to Na.}.  $E_c$ experiences an annual modulation, with a maximum in
January and a minimum in July.
Since iodine is much heavier than sodium, WIMP
interactions with iodine dominate the count rate (except for a high energy
tail above $E_c$ of iodine).  For WIMP masses heavier than 50 GeV,
the step lies above the DAMA threshold energy during peak months, and
can in principle be detected.

\begin{figure}
\includegraphics[width=0.45\textwidth]{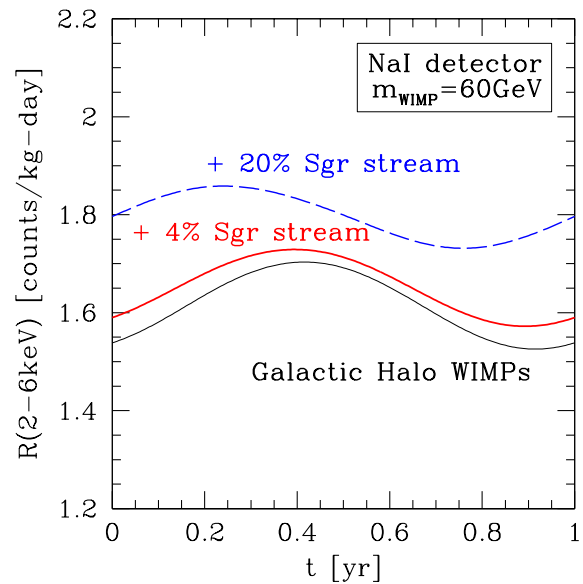}
\caption{
  Count rate of 60 GeV WIMPs over the course of a year in a NaI
  detector such as DAMA due to the halo alone as well as the sum of
  the halo and stream, for our reference direction for stream
  densities that are 4\% and 20\% of the halo, and in the 2-6 keVee
  bin.  }

\end{figure}

Figure 1 shows the time series of the count rate of 60 GeV WIMPs in
the detector,
integrated over the 2-6 keVee energy bin for which DAMA published results.
For a 4\% stream
contribution, the count rate of the halo alone peaks
on June 2, the stream alone on Jan.\ 15, and the sum of the two
on May 25.  The DAMA experiment finds that the count
rate in their data peaks on May 21 $\pm$ 22 days (1$\sigma$ error bars
\cite{DAMA}). 
The error band includes
June 2 (predicted by an isothermal halo model) but is
better fit by the presence of the stream. The peak date of
the total count rate (including stream) is extremely sensitive to the
direction and density of the stream.  For 60 GeV WIMPs and
a 20\% stream contribution, the peak date of the total signal in the 
2-6 keVee bin is March 30, in disagreement with the data. 
Thus DAMA cannot be seeing a stream with these parameters.

 \begin{figure}
\includegraphics[width=0.45\textwidth]{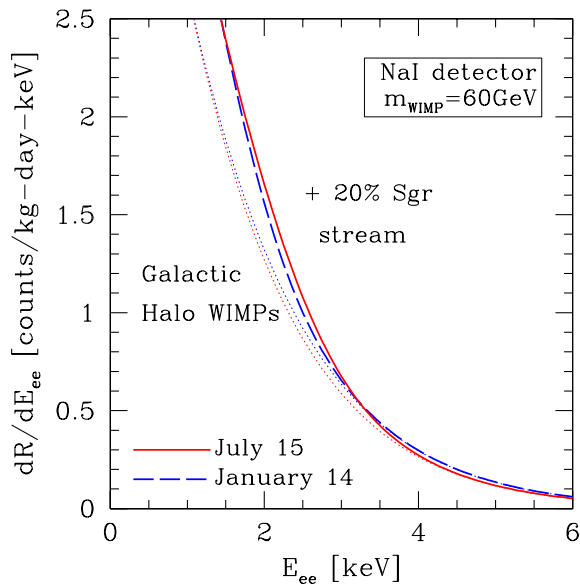}
\caption{Count rate of 60 GeV WIMPs in a NaI detector such as DAMA vs.
  recoil energy.  The dotted lines
  show the count rate from Galactic (isothermal) halo WIMPs alone.  The 
  solid and dashed lines show the step in
  the count rate if we include the Sgr stream WIMPS.  The plot 
  assumes that the stream contributes an
  additional 20\% of the local Galactic halo density, and comes from our 
  reference direction.  The solid and
  dashed lines are for July 15 and Jan.\ 14 respectively, the dates of
  maximum and minimum count rate for the stream.  }

\end{figure}

Figure 2 shows the count rate of 60 GeV WIMPs in a NaI detector such
as DAMA as a function of recoil energy.  A stream WIMP fraction of
$20\%$ is depicted to illustrate the step, though a $4\%$ WIMP
fraction is more likely.  The DAMA experiment has an enormous exposure
of 107731 kg-days.  For a 60 GeV WIMP mass and a Sgr stream in the
reference direction that is 4\% of the local halo density, we compute
the following number of events expected in the (2-3,3-4)keVee bins,
averaged over a year: (112372, 50894) with the stream and (108544,
50425) without the stream (assuming a detector resolution much smaller
than the bin size).  The location of $E_{c,max}$ due to iodine is
2.5keVee.  Thus the stream is detectable at the 11$\sigma$ level
($(112372-108544)/\sqrt{112372} = 11\sigma$) in the 2-3 keVee bin and
at the 2$\sigma$ level in the 3-4 keVee bin.
For higher masses, the value of $E_c$ is larger so that stream WIMPs
contribute out to higher energy recoils.  Stream WIMPs with higher
mass are detectable in the first few energy bins, even for rather low
dark matter densities in the stream.  For example, for 70 GeV WIMPs, a
4\% stream is detectable at the 16$\sigma$ level in the 2-3 keVee bin
and at the 5$\sigma$ level in the 3-4 keVee bin.  For 85 GeV WIMPs,
even a 2\%(1\%) stream is detectable at the 9$\sigma$ (5$\sigma$)
level in the 2-3 keVee bin and at the 5$\sigma$ (3$\sigma$) level in
the 3-4 keVee bin.

Experiments may be able to see the annual modulation of the step,
which is important to proving that they have indeed seen the Sgr
stream.  With the current DAMA energy resolution of 7.5\%, DAMA may be
able to make this identification in some cases but not in others.  For
the reference case, the energy resolution is better than the
difference in $E_c$ due to annual modulation (for those cases where
$E_c$ is above threshold).  Finding the Sgr modulation would be
persuasive in the interpretation of the observed annual modulation in
DAMA as indeed due to WIMPs.

{\it Conclusions:} Recent observations of the Sgr dwarf
spheroidal galaxy indicate the existence of tidal streams 
 that pass through the solar neighborhood.  It is possible that
stars associated with the stream have already been detected in the
solar neighborhood \cite{helmi}.  
If dark matter consists of WIMPs, the extra
contribution from the stream gives rise to a step-like feature in the
energy recoil spectrum in direct dark matter detection.  The location
of the step experiences an annual modulation that will be useful in
identifying the existence of the stream.  The count rate in the
detector due to stream WIMPs is also modulated annually. With 
our best estimates, the
maximum is on Jan.\ 15 and the minimum on July 16 for energy recoils near the
characteristic energy; for lower energy recoils, the phase is
opposite.  

For current and upcoming experiments, the step may lie above the
threshold energy and can be detected. DAMA may have it in
their current data.  For a 60 GeV WIMP mass and a very reasonable
stream density that is 4\% of the local halo, the stream is detectable
at the 11$\sigma$ level in the 2-3 keVee energy bin;
then the total count rate peaks on May 25, in excellent agreement with
the data.  The existence of the Sgr stream in the DAMA data may shift
the best fit WIMP mass and cross section\footnote{Since the DAMA data
  are not public, we are compelled to leave the analysis to the
  experimentalists.}.  It is possible that the discrepancy between
DAMA and EDELWEISS, CDMS, or ZEPLIN-I data may be resolved if the
stream is correctly included.  We also note that other experiments,
e.g., CDMS, CRESST, DRIFT, ZEPLIN, and XENON, should also be able to
see the stream, as discussed in our longer paper \cite{longer}.

K. F. and P. G. thank the DOE and the MCTP at the Univ.\ of Michigan
for support.  H. N. is supported by Research Corp.\ and the NSF (AST-0307571).

\end{document}